# Effect of Mindfulness and Mindful Art on Beginners and Experienced Meditators

Koonlin Eunice Chan
*Founder, Koonlin Mindfulness*
Shenzhen, China
430455759@qq.com

Joy Bose
*Global AI Accelerator, Ericsson*
Bangalore, India
joy.bose@ieee.org

*Abstract*—Mindfulness meditation has been proven to be effective in treating a range of mental and physical conditions. Mindful Art is a type of mindfulness meditation that comprises sessions of drawing, painting and sculpturing with mindfulness for a given length of time. To date, the efficacy of mindful art has not been systematically studied. In this paper, we describe an experimental pilot study on two groups of participants, a beginner group of 21 participants and an experienced meditation group of 9 participants, who had previously practiced mindfulness meditation for more than one year. The beginner group was instructed in mindfulness sitting and moving meditation, while the experienced group was instructed in mindful art making in addition to mindfulness meditation. The instructions were delivered remotely over Tencent Conference and WeChat. The sessions were of 90 minutes duration each, twice per week, with 45 minutes of home practice daily and the length of the study was 21 days. The blood pressure, pulse rate and breathing rates, as well as the subjective degree of relaxation were recorded at every session. At the end of the study, the experienced group reported higher average difference in breath rate and relaxation within each session, while the beginner group reported a greater degree of improvement in breath rate and relaxation over the period of the study, although their scores were lower on average than the experienced group.

## I. INTRODUCTION

Mindfulness has been defined as the practice of cultivating awareness that arises through paying attention, on purpose, in the present moment, non-judgmentally [1]. Mindfulness meditation is practiced in a variety of forms, including unstructured forms such as vipassana and mindfulness apps, and structured forms such as Mindfulness Based Cognitive Therapy (MBCT) and Mindfulness Based Stress Reduction (MBSR) [2]. Mindfulness has been found to be useful in treating a variety of mental and physical health conditions such as stress reduction, chronic pain [3], anxiety, and depression [4] and today is used in clinical settings in more than 50 countries around the world [5].

Mindfulness art [6] is a type of mindfulness practice involving practicing art mindfully. It has been used in a variety of forms such as mandala coloring books and free drawing. Mindfulness has also been successfully practiced in conjunction with art therapies [7]. Additionally, mindful viewing of relaxing images has been shown to have benefits for stress reduction among hospital patients [8]. Mindfulness art can especially appeal to people who are more visually or artistically oriented and want to practice mindfulness.

However, as of today there is a lack of structured mindfulness programs that incorporate mindfulness art. In this paper, we discuss a 3-week pilot study involving a structured program of mindfulness art along with general mindfulness, and its effects on biological stress indicators such as pulse rate, blood pressure and breathing rate. The subjects were a combination of beginners and experienced meditators based in China. The model of delivery of the mindfulness instruction was remotely via online conference and chat apps.

The rest of the paper is structured as follows: Section 2 discusses the study objectives and the design of the study, as well as how data from the study was collected and analyzed. Section 3 discusses the results of the data collection during the study, investigating the effect of the mindfulness study on blood pressure, pulse rate and breathing rate and the variation by age and gender. Section 4 concludes the paper and discusses avenues for future work.

## II. STUDY OBJECTIVE AND METHODOLOGY

### A. Study Objective

The objective of the pilot study was as follows: To understand the effect of general mindfulness meditation and mindfulness art meditation on a population of experienced and beginner meditators, undergoing mindfulness instruction remotely, with two 90-minute sessions per week, over a period of 3 weeks. This involves understanding the changes in biological indicators such as pulse rate, breathing rate and blood pressure, that may correlate to stress levels, as a result of mindfulness practice. The study seeks to compare the effect of mindfulness and mindfulness art on beginners as compared to more experienced meditators, as well as the variation of the effect depending on variables such as age group and gender.

### B. Study Design

The study conducted was an experimental pilot study. Before the study began, the participants had to sign a consent form indicating their consent for taking part in the study. Also, the personal data of the participants was anonymized before data collection: names were replaced with codes and age replaced with age range.

The participants were divided into the following two groups:

- Group A for experienced meditators who had previously undergone mindfulness meditation practice for over a year
- Group B for beginners to mindfulness meditation

Treatment period was 3 weeks, with daily practice and recording of heart rate and other data.

Group A participants were given instructions in General mindfulness as well as Mindfulness Art, while group B participants were given instruction only in General mindfulness meditation, comprising of sitting meditation and mindful physical exercises similar to chi gong.

The instruction period was about 90 minutes per session, with 2 sessions per week delivered remotely via Tencent conference and WeChat in the evening time around 8 pm in Beijing time zone. The participants were additionally given homework comprising 45 minutes of mindfulness practices to be done daily.

During the study, data related to biological parameters and other measures were collected from the participants. Both subjective measures (amount of relaxation) as well as objective measures (number of breaths per minute, pulse rate, high and low blood pressure) were considered.

### C. Content of a meditation session

A typical 90 minutes meditation session instruction for the participants, delivered remotely, consisted of the following stages:

The participants were first instructed to practice simple awareness and body scanning. After this, they were asked to focus on the "relaxation effect", to relax and meditate. The next part included mental noting and simple movement. The next part consisted of stress relief via simple physical and mental exercises. The next stage was to focus on emotions, particularly compassion. At this stage, the participants were asked for feedback and had to fill in a feedback form.

The next stage started with the participants reporting their blood pressure, pulse rate, breathing rate and relaxation. For the Group A participants, they engaged in tasks related to mindful drawing while paying attention to breathing (5 minutes), a free relaxation exercise (5 minutes), a body scan (15 minutes), followed by blood pressure, pulse and breathing rate measurement, followed by a session of mindful art drawing (5 minutes). Group B participants were instructed to mindfully eat raisins and drink water (15 minutes), followed by measurement of blood pressure, pulse rate and breathing rate (5 minutes).

Over the days, as the study progressed for 21 days, a number of other meditation techniques were included in the instruction. Some of the exercises were adapted from Mindfulness Based Stress Reduction (MBSR) program [2], a structured and widely used mindfulness 8-week program developed by Jon Kabat-Zinn, while others were adapted from Chi Gong based exercises. These included observing breathing by pressing the abdomen (15 minutes), chi gong exercises with raising arms and turning waist (5 minutes), three-step relaxation meditation (15 minutes), measuring blood pressure, pulse and breathing, breathing drawing (5 minutes). The main difference between group A and group B participants was a session on mindfulness art for group A lasting 15 minutes, and an equivalent session on body and eating mindfulness for group B also lasting 15 minutes.

### D. Data Collection

The following data was collected from the participants before, during and after the practice:

- High Blood Pressure (HBP)
- Low Blood Pressure (LBP)
- Pulse Rate
- Number of breaths counted in 60 seconds

A subjective measure called degree of relaxation (on a scale of 1 to 10, 10 is highest) was also collected before and after the practice.

### E. Data Analysis

For the participants of group A and group B, the average of the measures was taken for each week in the three weeks. The resulting graphs were plotted, to notice the following effects:

- Effect on blood pressure, pulse rate, breaths per minute and degree of relaxation within the meditation session
- Effect on blood pressure, pulse rate, breaths per minute and degree of relaxation over the 3-week course duration
- Comparison of the effect for experienced meditators vs beginners
- Comparison of the effect for male vs female meditators
- Comparison of the effect for different age groups of meditators

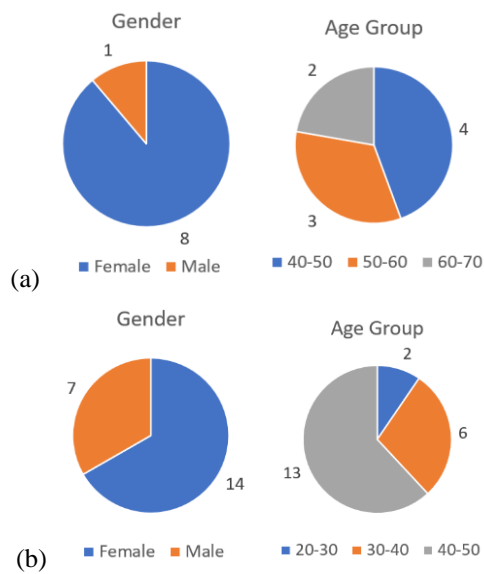

Fig. 1. Demographic profiles of (a) Group A and (b) Group B participants by gender and age group

### III. RESULTS

In this section, we present the results of the analysis of data collected from Group A and Group B participants during the study.

### A. Participant profiles

Fig. 1 shows the participant profiles for group A and group B. Group A mostly had only one male participant and 8 females, while the gender ratio was more even for group B, though still skewed 2:1. Group A consisted of older participants (aged between 40-70) while group B had younger participants (age range 20-50).

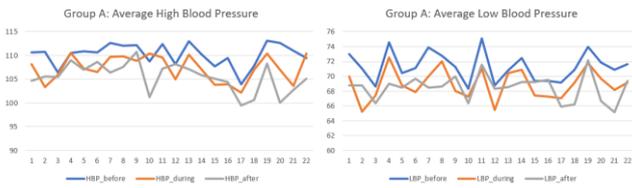

Fig. 2. Average High Blood Pressure (HBP) and Low Blood Pressure (LBP) before, during and after each session measured for Group A participants over the course of the 21 day pilot study

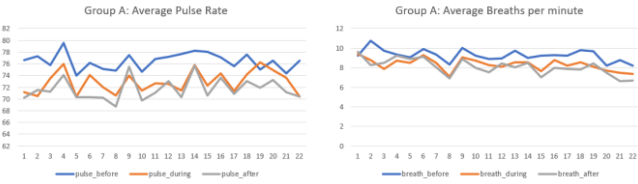

Fig. 3. Average Pulse Rate and Breathing Rate before, during and after each session measured for Group A participants over the course of the 21 day pilot study

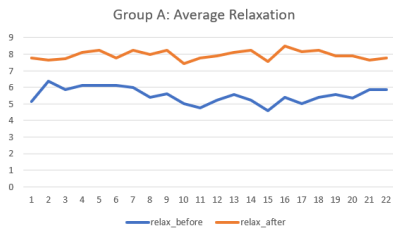

Fig. 4. Average Relaxation Level before and after each session measured for Group A participants over the course of the 21 day pilot study

### B. Average measurements of blood pressure, pulse rate, breathing rate and relaxation

The HBP, LBP, breathing rate and subjective degree of relaxation are measured before, during and after the study (only before and after for relaxation). The results for group A are plotted in figures 2-4 and for group B in figures 5-7.

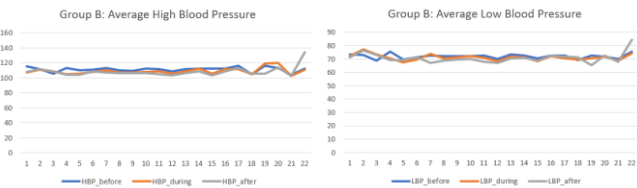

Fig. 5. Average High Blood Pressure (HBP) and Low Blood Pressure (LBP) before, during and after each session measured for Group B participants over the course of the 21 day pilot study

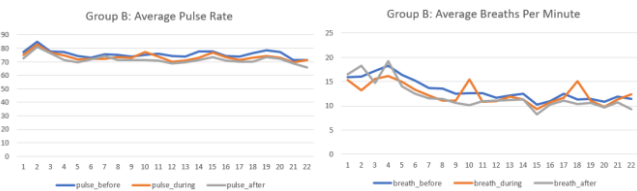

Fig. 6. Average Pulse Rate and Breathing Rate before, during and after each session measured for Group B participants over the course of the 21 day pilot study

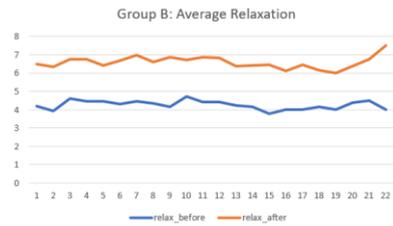

Fig. 7. Average Relaxation Level before and after each session measured for Group B participants over the course of the 21 day pilot study

Examining the graphs, we can see that Group B showed better improvement over the course of 21 days in average pulse rate and breaths per minute, while group A (experienced meditators) had better average scores (lower pulse rate and breath rate and relaxation) than group B but did not show too much improvement over the study period. The average pulse rate for group A students within each meditation session showed a bigger difference consistently than for group B.

### C. Some other observations

Some other observations from the plotted charts included the following:

- Many Group B participants did not fill data regularly. Group A generally filled out the data regularly. However, this effect got averaged out since we took the averages for the data.

- For Group A, Pulse and breaths slowed more after meditation. For Group B, No obvious trend in Pulse, Breath, however Group B participants showed improvement over time.

- For HBP and LBP: No obvious trend in HBP and LBP in either group, although group A slightly slowed down.

- Relaxation: Both Group A and Group B observed higher subjective measures of relaxation after the meditation sessions.

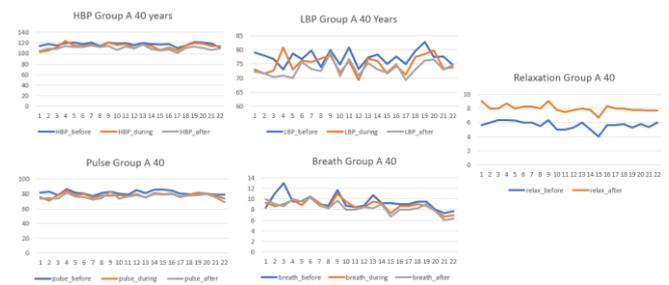

Fig. 8. HBP, LBP, breathing rate and subjective degree of relaxation for Group A participants aged 40-50 years

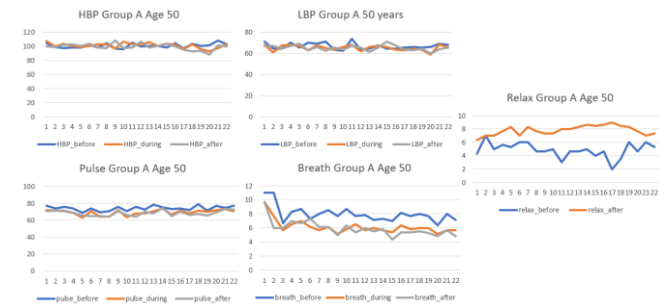

Fig. 9. HBP, LBP, breathing rate and subjective degree of relaxation for Group A participants aged 50-60 years

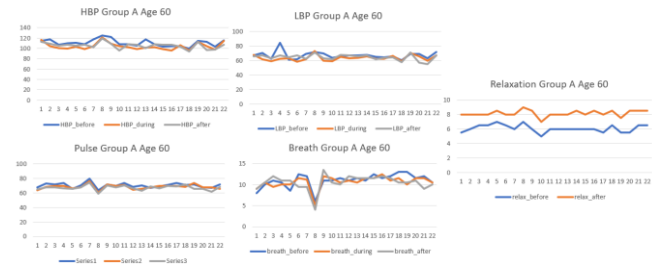

Fig. 10. HBP, LBP, breathing rate and subjective degree of relaxation for Group A participants aged 60-70 years

*D. Affect of Age on group A participants*

Figures 8-10 present the measurements for group A participants at age range 40-50, 50-60 and 60-70. The variation related to age was generally less pronounced for group A participants. The 50-60 years group A participants showed better improvement (graph sloping down) than other age groups in breathing rate over the course of the 21 days of the study. The average HBP and LBP numbers show a variation across different age ranges for group A.

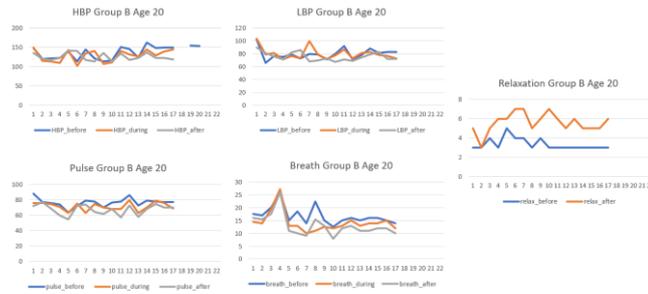

Fig. 11. HBP, LBP, breathing rate and subjective degree of relaxation for Group B participants aged 20-30 years

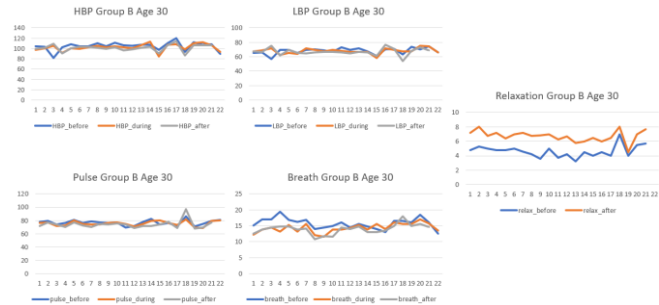

Fig. 12. HBP, LBP, breathing rate and subjective degree of relaxation for Group B participants aged 30-40 years

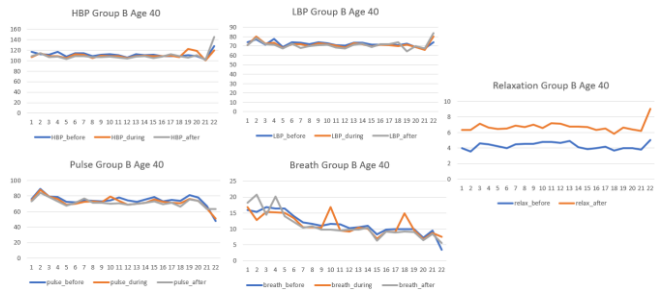

Fig. 13. HBP, LBP, breathing rate and subjective degree of relaxation for Group B participants aged 40-50 years

*E. Affect of Age on Group B participants*

Figures 11-13 present the measurements for group B participants at age range 20-30, 30-40 and 40-50. The variation related to age was generally more pronounced for group B than for group A. In particular, the 20-30 age participants did not fill up most of the data in the last few days of the study, indicating a higher dropout rate. Also, the data variability was higher for younger age group participants in group B. The 40-50 years group B participants showed the best and most consistent improvement (graph sloping down) in breathing rate and pulse rate over the course of the 21 days of the study. The average HBP and LBP numbers also show a variation across different age ranges.

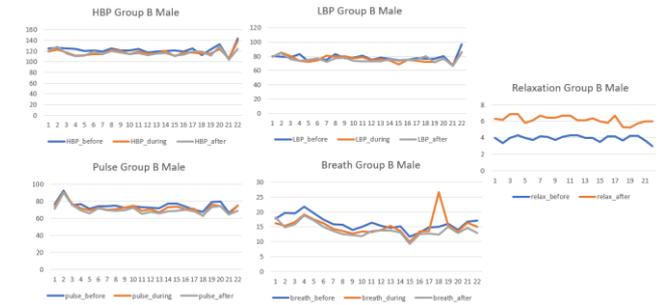

Fig. 14. HBP, LBP, breathing rate and subjective degree of relaxation for male Group B participants

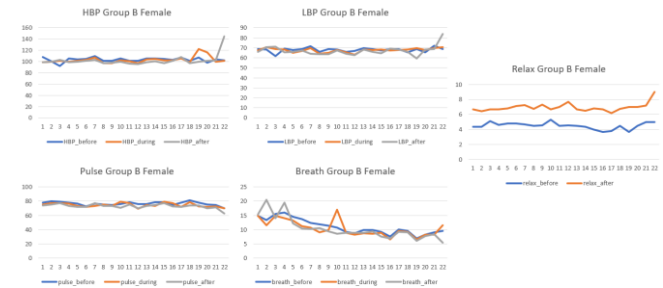

Fig. 15. HBP, LBP, breathing rate and subjective degree of relaxation for female Group B participants

*F. Affect of Gender on group B participants*

Figures 14-15 present the measurements for male and female group B participants. Since group A was heavily skewed in favor of the female gender, we did not plot the male vs female data for group A. The average HBP and LBP values are different from males as compared to females, although this may be unrelated to the meditation. Females showed a better improvement in breath rate over the 21 days duration< while the graph for males was more variable.

## IV. CONCLUSION AND FUTURE WORK

In this paper, we have presented the results of a pilot study conducted remotely over WeChat and Tencent conference for 30 people divided into two groups, with the first group A having 9 experienced meditators and mindfulness art as one of the mindfulness components, while the second group B had 21 beginner meditators and no mindfulness art. We have studied and recorded various effects related to gender and age range of the participants. In general, we found that the beginner meditators showed a better improvement over the course of the 21-day study while experienced meditators showed better scores on average.

Generally, structured mindfulness studies such as MBSRs are for 8-weeks, while our pilot study was only for 3 weeks. In future we plan to do a longer study and record a few additional parameters including subjective degree of mindfulness.

ACKNOWLEDGMENT

The authors are very grateful to Dr. Ma, Frank (Liu Wei) and Helen (Feiyang) for their help throughout the study in the area of data collection and in conducting the study.